\documentclass[12pt]{article}

\usepackage{hyperref}%
\hypersetup{%
   breaklinks,%
   ocgcolorlinks,
   colorlinks=true,%
   urlcolor=[rgb]{0.0,0.0,0.5},%
   linkcolor=[rgb]{0.45,0.0,0.0},%
   citecolor=[rgb]{0,0,0.45}
}
\usepackage{color,graphicx,latexsym}%
\usepackage{amsmath,amssymb}%
\usepackage{pifont}%
\usepackage{xspace}%
\usepackage[cm]{fullpage}

\newtheorem{theorem}{Theorem}[section]

\newcommand{\etal}{\textit{et~al.}\xspace}

\newcommand{\eps}{{\varepsilon}}
\newcommand{\reals}{{\mathbb R}}

\providecommand{\Matousek}{Matou{\v s}ek\xspace}
\providecommand{\Komlos}{Koml\'os\xspace}

\newcounter{newx}
\newcommand{\atgen}{\symbol{'100}}
\newcommand{\SarielThanks}[1]{\thanks{Department of Computer Science;
      University of Illinois; 201 N. Goodwin Avenue; Urbana, IL,
      61801, USA; {\tt sariel\atgen{}illinois.edu}.%
   } #1}

\def\A{{\cal A}}

\def\F{{\cal F}}
\def\G{{\cal G}}
\newcommand{\CalH}{{\cal H}}
\def\dg{{\sf deg}}
\def\bd{{\partial}}

\def\R{{\cal R}}

\newcommand{\cardin}[1]{\left|#1\right|}

\newcommand{\si}[1]{#1}

\renewcommand{\th}{th\xspace}
\providecommand{\Erdos}{Erd{\H{o}}s\xspace}

\begin{document}

\title{$\eps$-Nets for Halfspaces Revisited%
   \thanks{%
      Work on this paper by Sariel Har-Peled~was partially supported
      by an NSF CAREER award CCR-0132901.  Work by Haim Kaplan was
      partially supported by Grant 975/06 from the Israel Science
      Fund. Work by Micha Sharir was partially supported by NSF Grant
      CCF-05-14079, by a grant from the U.S.-Israeli Binational
      Science Foundation, by grant 155/05 from the Israel Science
      Fund, Israeli Academy of Sciences, by a grant from the
      \si{AFIRST} joint French-Israeli program, and by the Hermann
      Minkowski--MINERVA Center for Geometry at Tel Aviv University. Work by Shakhar Smorodinsky was
      partially supported by Grant 1136/12 from the Israel Science
      Fund.%
   }%
}

\author{
   Sariel Har-Peled\SarielThanks{}%
   \and%
   Haim Kaplan\thanks{%
      School of Computer Science, Tel Aviv University, Tel Aviv 69978,
      Israel; \texttt{haimk@tau.ac.il} }%
   \and%
   Micha Sharir\thanks{%
      School of Computer Science, Tel Aviv University, Tel Aviv 69978,
      Israel, and Courant Institute of Mathematical Sciences, New York
      University, New York, NY 10012, USA; \texttt{michas@tau.ac.il} }
   \and%
   Shakhar Smorodinsky\thanks{%
      Department of Mathematics, Ben Gurion University of the Negev,
      Be'er Sheva 84105, Israel; \texttt{shakhar@math.bgu.ac.il} } }

\date{\today}

\maketitle

\begin{center}
    \begin{minipage}[c]{5.0in}
        \footnotesize
        \begin{quotation}
            ``It is a damn poor mind indeed which can't think of
            at least two ways to spell any word.''

            \qquad\qquad\qquad -- \textrm{Andrew Jackson}
        \end{quotation}

        ~

        ~

    \end{minipage}
\end{center}

\begin{abstract}
    Given a set $P$ of $n$ points in $\reals^3$, we show that, for any
    $\eps >0$, there exists an $\eps$-net of $P$ for halfspace ranges,
    of size $O(1/\eps)$. We give five proofs of this result, which are
    arguably simpler than previous proofs \cite{msw-hnlls-90,
       cv-iaags-07, pr-nepen-08}.  We also consider several related
    variants of this result, including the case of points and
    pseudo-disks in the plane.
\end{abstract}


\section{Introduction}

Since their introduction in 1987 by Haussler and
Welzl~\cite{hw-ensrq-87} (see also Clarkson~\cite{c-narsc-87} and
Clarkson and Shor~\cite{cs-arscg-89} for related concepts),
$\eps$-nets have become one of the central concepts in computational
and combinatorial geometry, and have been used in a variety of
applications, such as range searching, geometric partitions, and
bounds on curve-point incidences, to name a few. We recall their
definition: A {\em range space} $(X,\R)$ is a pair consisting of an
underlying universe $X$ of objects, and a certain collection $\R$ of
subsets ({\em ranges}) of $X$. Of particular interest are range spaces
of {\em finite VC-dimension}; skipping the exact definition, it
suffices to require that, for any finite subset $P\subset X$, the
number of distinct sets $r\cap P$, for $r\in\R$, is $O(|P|^d)$, for
some constant $d$ (which is upper bounded by the VC-dimension of
$(X,\R)$).

Given a range space $(X,\R)$, a finite subset $P\subset X$, and a
parameter $0<\eps<1$, an {\em $\eps$-net} for $P$ (and $\R$) is a
subset $N\subseteq P$ with the property that any range $r\in\R$ with
$|r\cap P|\ge \eps|P|$ contains an element of $N$. In other words, $N$
is a hitting set for all the ``heavy'' ranges.

The important result of Haussler and Welzl asserts that, for any
$(X,\R)$, $P$, and $\eps$ as above, such that $(X,\R)$ has finite
VC-dimension $d$, there exists an $\eps$-net $N$ of size
$O\left(\frac{d}{\eps}\log\frac{1}{\eps}\right)$, and that a random
sample of $P$ of that size is an $\eps$-net with constant
probability. In particular, the size of $N$ is independent of the size
of $P$.

In geometric applications, this abstract framework is used as
follows. The ground set $X$ is typically a set of simple geometric
objects (points, lines, hyperplanes), and the ranges in $\R$ are
defined as a Boolean combination of intersection with (or, for point
objects, containment in) simply-shaped regions (halfspaces, balls,
simplices, etc.), formally required to be regions of {\em constant
   descriptive complexity}, meaning that they are semi-algebraic sets
defined in terms of a constant number of polynomial equations and
inequalities of constant maximum degree.  It is known that in such
cases the resulting range space $(X,\R)$ has finite VC-dimension (see,
e.g., \cite{m-ldg-02}).

One of the major questions in the theory of $\eps$-nets, posed since
their introduction more than 25 years ago, is whether the factor
$\log\frac{1}{\eps}$ in the upper bound on their size is really
necessary, especially in ``normal'' (geometric) situations. To be
precise, it has been known, pretty early in the game, that in the
general abstract context the answer is ``yes''. This was shown by
\Komlos, Pach and Woeginger~\cite{kpw-atben-92} back in 1992, using a
randomized construction on abstract hypergraphs (see also
\cite{pa-cg-95}).

Concerning simple geometric range spaces, Alon~\cite{a-nllbp-12} was
the first to obtain a ``non-linear'' lower bound (i.e., more than 
$C/\eps$ for any positive constant $C$) on the size of $\eps$-nets 
for the range space of points and triangles in the plane. More recently, 
Pach and Tardos \cite{pt-tlbse-13} showed a lower bound of 
$\Omega(\tfrac{1}{\eps}\log \log \tfrac{1}{\eps})$ for points and axis-parallel
rectangles in the plane, and, more significantly, a lower bound
$\Omega( \tfrac{1}{\eps}\log \tfrac{1}{\eps})$ for the dual range space 
(namely, a range space where $X$ is a collection of axis-parallel rectangles, 
and each range is the subset of the rectangles that contains some given point).
A simple reduction then leads to a lower bound of 
$\Omega( \tfrac{1}{\eps}\log \tfrac{1}{\eps})$ on the size 
of $\eps$-nets for points and halfspaces in $\reals^4$. 
In other words, $\eps$-nets of size smaller than
$\Theta( \tfrac{1}{\eps}\log \tfrac{1}{\eps})$, even in simple geometric
contexts, seem to be a relatively rare phenomenon.

Nevertheless, many improved upper bounds on the size of $\eps$-nets in
a variety of geometric contexts have been
obtained~\cite{aes-ssena-10, cv-iaags-07, kk-iagsg-11, pr-nepen-08}. 
Some of these bounds are linear (in $1/\eps$), while
others are in between linear and the general bound
$O\left(\frac{1}{\eps}\log\frac{1}{\eps}\right)$.  The first linear
bound on the size of $\eps$-nets has been obtained by \Matousek,
Seidel and Welzl~\cite{msw-hnlls-90}, 25 years ago, for the special
cases of points and halfspaces in two and three dimensions, and for
some other related special cases. The proofs in \cite{msw-hnlls-90}
are somewhat involved, and appear to have some technical difficulties,
which are corrected in a revised version.  \Matousek has given an
alternative construction for halfspaces (in two and three dimensions),
with the same bounds, using his shallow-cutting lemma \cite{m-rph-92}.
Additional progress was made more recently.  Clarkson and
Varadarajan~\cite{cv-iaags-07}, essentially adapting \Matousek's
technique to their more general setting, have come up with a technique
for constructing small-size $\eps$-nets in geometric dual range spaces,
where, as above, the objects of $X$ are simply shaped regions, and each 
range is the subset of regions that are stabbed by some point.
If the combinatorial complexity of the union of any finite number $r$ 
of the regions is small, specifically $o(r\log r)$, then the corresponding
range space admits $\eps$-nets of size $o\left(\frac{1}{\eps}\log\frac{1}{\eps}\right)$.
Pyrga and Ray~\cite{pr-nepen-08} have proposed a general abstract
scheme for constructing small-size $\eps$-nets in hypergraphs (that
is, range spaces) which satisfy certain properties, and have applied
it to the special cases of halfspaces in two and three dimensions, and
to a few other related instances. Later, Aronov, Ezra and
Sharir~\cite{aes-ssena-10} have shown the existence of $\eps$-nets of
size $O\left(\frac{1}{\eps}\log\log\frac{1}{\eps}\right)$ for the
range space of points and axis-parallel rectangles in the plane 
(a tight bound as subsequently shown by Pach and Tardos), as well as 
improved bounds for the size of $\eps$-nets for several other range spaces. 
See also King and Kirkpatrick~\cite{kk-iagsg-11} for another recent improved 
bound for $\eps$-nets for range spaces related to art-gallery visibility.

\paragraph{Our results.}
In this note we re-examine the techniques in 
\cite{ cv-iaags-07, m-rph-92, pr-nepen-08}, extract from them the main 
ingredients, and dress them up in different and arguably simpler proofs,\footnote{%
   Obviously, simplicity is in the eye of the beholder, and we can
   only hope that the reader share our feeling that the proofs are
   indeed simpler.}  
which exploit the geometry of arrangements of planes in 3-space 
and of several other geometric structures. One of
our (more complicated) proofs is essentially a restatement of
\Matousek's proof in \cite{m-rph-92}, and of the similar (more
general) proof in Clarkson and Varadarajan~\cite{cv-iaags-07}; 
we give it here for the sake of
completeness. The other four proofs are (in our opinion) considerably
simpler. They use as a key ingredient a construction of Pyrga and
Ray~\cite{pr-nepen-08}, but the analyses of the size of the resulting
structure are different.

It is our hope that the multitude
of proofs will shed more light on the problem, and that some of them
might eventually be extended to other geometric range spaces.


Finally, we note that, for the case of halfspaces in three (and two)
dimensions, one can construct linear-size $\eps$-nets in an efficient
deterministic manner, by a careful implementation of the construction
given in the proofs, using standard techniques due to
\Matousek~\cite{m-dcg-96}.


\section{Small-size $\eps$-nets for halfspaces in $\reals^3$}

Let $P$ be a set of $n$ points in $\reals^3$ in general position, and
let $\CalH$ be the family of all (closed) halfspaces (bounded by
planes).  The main result of this section is:
\begin{theorem}%
    \label{main3d}%
    Given a set $P$ of $n$ points in $\reals^3$ in general position,
    and a parameter $0<\eps\le 1$, there exists an $\eps$-net for
    $(P,\CalH)$ of size $O(1/\eps)$.
\end{theorem}

\noindent{\bf Remark.}
Clearly, the theorem implies that a similar result holds for the range
space of points and halfplanes in the plane. This, however, can be
established using considerably simpler and shorter proofs, see, e.g.,
\cite{sy-pchp-12, w-enh-89}. The general position assumption in
the theorem is for the sake of simplicity of exposition, and was used
also in the previous work \cite{msw-hnlls-90}.

\medskip

We give five different proofs of the theorem. The first four use the
same construction, inspired by the approach of Pyrga and
Ray~\cite{pr-nepen-08}, but differ in the way they show that the resulting
$\eps$-net has small size. The fifth proof is essentially a
restatement of the proof of \Matousek \cite{m-rph-92}, and of the
similar more general proof in \cite{cv-iaags-07}, and is included here
for the sake of completeness; it is (in our opinion) somewhat more
complex than the first four proofs.

\paragraph{The construction.}
Without loss of generality, it suffices to construct an $\eps$-net
for {\em lower} halfspaces. A symmetric construction will yield an
$\eps$-net for {\em upper} halfspaces, and the union of the two nets
will be an $\eps$-net for all halfspaces.
Let $\CalH^-$ denote the set of all lower halfspaces.

We fix some constant fraction $0<\beta<1$ (different choices of
$\beta$ will be made in the different proofs), and construct a
maximal collection $\F$ of lower halfspaces with the following
properties (where we assume that $\eps\le 1/2$, for otherwise
there exists a constant-size $\eps$-net, by the general theory
in \cite{hw-ensrq-87}):

\begin{description}
    \item{(a)} Each halfspace in $\F$ contains between $\eps n$ and
    $2\eps n$ points of $P$.

    \item{(b)} For any pair of distinct halfspaces $h,g\in\F$,
    we have $|h\cap g\cap P| \le \beta\eps n$.
\end{description}

It is easily seen that $\F$ is finite.

For each halfspace $h\in\F$, we construct a $(\beta/2)$-net $N_h$ for
the set system $(h\cap P,\CalH^-)$, of size
$O\left(\frac{1}{\beta}\log\frac{1}{\beta}\right) = O(1)$, using the
standard bounds on the size of $\eps$-nets~\cite{hw-ensrq-87}, and
form the union $N^{(1)}=\bigcup_{h\in\F} N_h$.

We next repeat the same construction for each value
$\eps_j = 2^{j-1}\eps$, for $j=1,2,\ldots$, using the same parameter
$\beta$ for each $\eps_j$. We obtain a sequence of subsets
$N^{(j)}\subseteq P$, and we set $N$ to be their union.
Let us also denote by $\F^{(j)}$ the maximal set of halfspaces
constructed at the $j$\th step.

\paragraph{$N$ is an $\eps$-net.}
It is easy to see that $N$ is an $\eps$-net for $(P,\CalH^-)$.
Indeed, let $h$ be a lower halfspace which contains at least $\eps n$
points of $P$. There exists $j\ge 1$ such that $2^{j-1}\eps n\le
|h\cap P| < 2^j\eps n$.  If $h\in\F^{(j)}$ then it certainly contains
a point of $N$ (it contains the nonempty subset $N_h\subseteq N$).
Otherwise, $\F^{(j)}\cup\{h\}$ must violate property (b), so
$\F^{(j)}$ contains a lower halfspace $g$ such that
$$
|h\cap g\cap P| > \beta\eps_j n = \beta 2^{j-1}\eps n \ge
\frac{\beta}{2} |g\cap P| .
$$
Hence, by construction, $h$ must contain a point of $N_g$,
and thus of $N$.

The main challenge is to argue that $|\F|=|\F^{(1)}| = O(1/\eps)$.
Since $\beta$ is a constant, this would imply that
$|N^{(1)}|=O(1/\eps)$ too, and, more generally, that
$|N^{(j)}|=O(1/(2^{j-1}\eps))$. Summing these bounds, we would
then obtain $|N|=O(1/\eps)$.

\medskip

\noindent{\bf Remark.}
The approach just presented, which uses a geometric progression of
values of $\eps$ and a corresponding sequence of constructions,
appears to be unnecessarily over-complicated.  Specifically, for
halfspaces, a single step suffices: We only construct $N^{(1)}$, and
claim that it is an $\eps$-net. Indeed, if $h$ is any lower halfspace
containing at least $\eps n$ points of $P$ then, by the general position 
assumption, we can shrink it to
another halfspace $h'\subseteq h$ which contains exactly $\eps n$
points of $P$. The preceding argument implies that $h'$ contains a
point of $N^{(1)}$, and therefore so does $h$. The reason for
complicating the construction is that it can also handle ranges which
do not have this ``shrinking property'', namely the property that any
range can be shrunk to a smaller range which contains any prescribed
number of points of $P$.  The same trick of repeatedly doubling $\eps$
has also been used by Pyrga and Ray~\cite{pr-nepen-08}.

\medskip

We give four proofs of the claim that $|\F|=O(1/\eps)$.
Each proof is based on a specific (and different) choice of $\beta$,
which is spelled out during the analysis.

\paragraph{First proof.}
For each $h\in\F$ let $\pi_h$ denote its bounding plane.
By slightly perturbing these planes, without changing any of the
subsets $h\cap P$, for $h\in\F$, we may assume that the planes $\pi_h$
are in general position. We claim
that all the planes $\pi_h$ appear on their upper envelope
$E$. Indeed, suppose to the contrary that there exists $h\in\F$ such
that $\pi_h$ lies fully below the envelope.
Let $v$ be the vertex of the envelope closest to $h$.
Clearly, the union of the three halfspaces $h_1,h_2,h_3\in\F$
defining $v$ cover $h$; that is, $h\subseteq h_1\cup h_2\cup h_3$.
Hence, for at least one index $i\in\{1,2,3\}$, we have
$|h\cap h_i\cap P| \ge \frac13 |h\cap P| \ge \frac13 \eps n$,
which contradicts property (b) if we choose $\beta < 1/3$.

\begin{figure}[htb]
    \begin{center}
\begin{picture}(0,0)%
\includegraphics{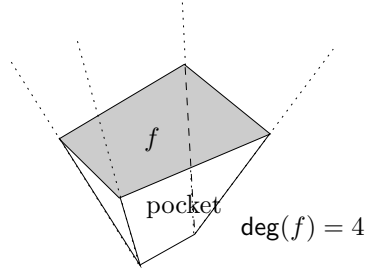}%
\end{picture}%
\setlength{\unitlength}{3158sp}%
\begingroup\makeatletter\ifx\SetFigFont\undefined%
\gdef\SetFigFont#1#2#3#4#5{%
  \reset@font\fontsize{#1}{#2pt}%
  \fontfamily{#3}\fontseries{#4}\fontshape{#5}%
  \selectfont}%
\fi\endgroup%
\begin{picture}(2574,2094)(1834,-2090)
\put(2910,-1141){\makebox(0,0)[lb]{\smash{{\SetFigFont{10}{12.0}{\rmdefault}{\mddefault}{\updefault}{\color[rgb]{0,0,0}$f$}%
}}}}
\put(2926,-1665){\makebox(0,0)[lb]{\smash{{\SetFigFont{10}{12.0}{\rmdefault}{\mddefault}{\updefault}{\color[rgb]{0,0,0}pocket}%
}}}}
\put(3669,-1839){\makebox(0,0)[lb]{\smash{{\SetFigFont{10}{12.0}{\rmdefault}{\mddefault}{\updefault}{\color[rgb]{0,0,0}$\dg(f)=4$}%
}}}}
\end{picture}%

        \caption{A pocket.}
        \label{fig:poc}
    \end{center}
\end{figure}

Put $t=|\F|$, and consider $E$ as a planar map, which has $t$ faces.
Define the {\em degree} $\dg(f)$ of a face $f$ of $E$, lying on some
plane $\pi_h$, to be the number of planes $\pi_g$ which appear on
the $1$-level of $\A(\F)$ directly below $f$ (see Figure~\ref{fig:poc}
for an illustration). In general, each such plane $\pi_g$ either meets
$\bd f$ or contributes a face to the $1$-level which lies fully
below $f$; the second case is impossible,
though, for then $\pi_g$ would not appear on the upper envelope.
Hence, assuming general position,
$\dg(f)$ is equal to the number of edges of $f$.

By Euler's formula, the number of edges of the upper envelope
of $t$ planes in $\reals^3$ is at most $3t-6$.
Since each such edge participates in two pockets, we get
$\sum_f \dg(f) < 6t$, where the sum extends over all faces $f$ of $E$.
Hence, at least half of the $t$ faces of $E$ have degree at most $11$;
refer to these faces as \emph{light}.

Let $f$ be one of these faces, and let $h$ be the corresponding
halfspace. Then we have
\begin{align*}
    h\setminus \left ( \cup\F\setminus\{h\} \bigl. \right)%
    =%
    h \setminus \cup \, \F_h ,
\end{align*}
where $\F_h$ is the set of all halfspaces which contribute to the
degree of $f$; refer to this expression as the {\em pocket} of $f$ (or
of $h$). By the choice of $f$, we have $|\F_h|\le 11$.
Hence, the number of points of $P$ in the pocket is at least
$$
|h\cap P| - \sum_{g\in\F_h} |g\cap h\cap P| \ge
\eps n - 11\beta\eps n \ge \frac12\eps n ,
$$
if we choose $\beta\le 1/22$.

Since the pockets are clearly pairwise disjoint, the overall number of
points of $P$ in the pockets of the at least $t/2$ light faces
is at least $\frac14t\eps n$. Hence we have $\frac14t\eps n \le n$,
implying that $t\le 4/\eps$, as claimed.  $\Box$

\paragraph{Second proof.}
Here, in an attempt to generalize the analysis to other kinds of range
spaces, we do not assume that all the bounding planes appear on the
upper envelope. Let $\F_0$ denote the subset of those halfspaces in
$\F$ whose bounding planes do appear on the envelope of $\F$, and put
$t_0=|\F_0|$. For $i \geq 1$, let $\F_i$ be the subset of halfspaces
in $\G_i =\F \setminus (\F_0 \cup \cdots \cup \F_{i-1})$ whose bounding
planes appear on the upper envelope of $\G_i$. We repeat this peeling
process, so at the last stage $k$, all of $\F$ is exhausted, and
$\F_{k+1}=\emptyset$.

The overall complexity of the $0$- and $1$-levels of $\A(\G_i)$ is at
most $c(t_{i}+t_{i+1})$, where $t_i = \cardin{\F_i}$, and where $c$ is
some absolute constant. This follows from Euler's formula for the
$0$-level (as in the first proof), combined with the random sampling
technique of Clarkson and Shor~\cite{cs-arscg-89} (or, alternatively,
the simpler scheme of Tagansky~\cite{t-ntasa-96}).  For this, note
that any plane that shows up on the $0$- or $1$-level in $\A(\G_i)$
must belong to $\F_i\cup\F_{i+1}$.  That is, every time we peel away
the halfspaces touching the envelope, we also remove the pockets
touching the envelope, ``exposing'' their lower boundaries to the
upper unbounded cell, and making these boundaries participate in the
next envelope.  Recall that here we no longer assume that all the
planes bounding a pocket reach the boundary of its upper facet, as in
the preceding proof.  After the removal of $\F_i$, the new pockets of
the bounding planes of halfspaces in $\F_{i+1}$ are (pairwise disjoint
and) disjoint from the pockets of the preceding stages.

Summing over all stages, we obtain a total of $t=t_0+t_1+\cdots+t_k$
pockets (each plane contributes exactly one pocket), and the sum of
the degrees of the corresponding halfspaces (which we bound by the
complexity of their pockets) is at most
$c((t_0+t_1)+(t_1+t_2)+\cdots+(t_{k-1}+t_k)+(t_k))\le 2ct$.
Hence, the average degree of a halfspace is at most $2c$,
and therefore at least half of the pockets are of halfspaces with
degree at most $4c$. Hence, choosing $\beta=1/(8c)$, and arguing
as above, the bound $t=O(1/\eps)$ follows.
$\Box$

\paragraph{Third proof.}
The preceding proof peels off the halfspaces of $\F$ in batches, where
in each step we remove many halfspaces. An alternative, and perhaps
somewhat more natural approach is to peel them off one by one.
In this proof we use a random peeling order, and exploit known properties
of randomized incremental constructions of upper envelopes to
establish the desired bound on the size of $\F$.

Specifically, draw a random permutation of $\F$, which we write as
$(h_1,h_2,\ldots,h_t)$, and insert the halfspaces one by one in this
order.  When $h_j$ is inserted, it adds to the union a new (possibly
empty) pocket, which, as above, is $h_i\setminus\bigcup_{j<i} h_j$;
the pocket is nonempty if and only if $\pi_{h_j}$ appears on the
current upper envelope $E_j$. An obvious but crucial property is that
all these pockets are pairwise openly disjoint.  To facilitate the
analysis, we maintain a triangulation of $E_j$, using, e.g., bottom-vertex
triangulation of each facet, and maintain for each triangle $\tau$ a
{\em conflict list} of all the halfspaces $h$ not yet inserted, which
meet $\tau$. If $\tau$ lies fully in the interior of such a halfspace
$h$, then adding $h$ removes $\tau$ completely from the envelope;
otherwise, adding $h$ splits $\tau$ into a portion which is hidden
from the envelope and a portion which remains on the envelope.  The
remaining portions of the envelope are re-triangulated, the conflict
lists of the new triangles are computed from the lists of the
destroyed triangles, and the process continues. See, e.g.,
Seidel~\cite{s-barga-93} for a review of randomized incremental
constructions of this kind.

The standard analysis of randomized incremental constructions (see
\cite{s-barga-93}) implies that the expected overall number of
triangles that are ever created by the algorithm is at most $ct$, for
some absolute constant $c>0$.

Note that the complexity of the pocket of $h_j$, and thus the degree
of $h_j$, {\em at the time of its insertion}, is proportional to the
number of triangles of $E_{j-1}$ that are killed by $h_j$, either by
being fully eliminated, or by being split and replaced by new
triangles; refer to these degrees as the {\em degrees at birth}.
It follows that the expected sum of the degrees at birth of the
halfspaces is at most $c't$, for another absolute constant $c'>0$.
Hence the average degree at birth is at most $c'$, so at least
half of the halfspaces have degree at birth at most $2c'$.

Choose $\beta=1/(4c')$. Arguing as above, it is easy to see that the
number of points of $P$ in the pocket at birth
of such a `light' halfspace is
at least $\frac12\eps n$. Hence the total number of points of $P$ in
these pockets is at least $\frac14 t\eps n$. Since the pockets at
birth are pairwise disjoint, the bound on $t$ follows.
$\Box$

\paragraph{Fourth proof.}
This time we pass to the dual space, where each point $p\in P$ is
mapped to a dual plane $p^*$, and each range $h\in \F$ is mapped to a
dual point $h^*$ (actually, dual to the plane bounding $h$),
so that point $p$ lies in halfspace $h$ if and only
if the dual point $h^*$ lies above the dual plane $p^*$.  Let $P^*$
(resp., $\F^*$) denote the resulting set of $n$ dual planes (resp., of
$t$ dual points).  Note that the level of each point $h^*\in\F^*$ in
the arrangement $\A(P^*)$ is between $k=\eps n$ and $2k$, and that,
for any pair of distinct points $h^*, g^*\in\F^*$, the number of
planes that separate them is at least $2(1-\beta)k$.  Indeed, each of
these planes passes below exactly one of $h^*$, $g^*$.  Since at least
$k$ planes pass below $h^*$, at least $k$ planes pass below $g^*$, and
at most $\beta k$ planes pass below both of them, the claim follows.
Choose $r=(1-\beta)k$, and define, for each dual point $h^*\in\F^*$,
the ``ball'' $B_h$ of all the vertices of $\A(P^*)$ which are
separated from $h^*$ by fewer than $r$ planes.  As argued by
Welzl~\cite{w-stlcn-92}, the number of vertices in $B_h$ is
$\Omega(r^3)=\Omega(k^3)$. By construction, no vertex of $\A(P^*)$ can
appear in more than one of these balls. Indeed, let $d_{P^*}(u,v)$
denote the number of planes of $P^*$ which separate $u$ and $v$; then
$d_{P^*}$ satisfies the triangle inequality, from which the claim
follows readily. Moreover, since $h^*$ lies at level between $k$ and
$2k$, the level of any of these vertices is between $k-r$ and $2k+r$,
so they all lie at level at most $3k$.  As shown by Clarkson and
Shor~\cite{cs-arscg-89}, the overall number of vertices of $\A(P^*)$
at level at most $3k$ is $O(nk^2)$.  Hence the number $t$ of balls
satisfies $tk^3=O(nk^2)$, and thus $t=O(n/k)=O(1/\eps)$.  $\Box$

\paragraph{Fifth proof.}
This proof is an adaptation of \Matousek's proof on shallow cutting
\cite[Section 5]{m-rph-92}, and also of the proof in
\cite{cv-iaags-07}, which itself is an extension of \Matousek's proof;
we include it here for the sake of completeness.  In this proof we
abandon the set $\F$.  As in the preceding proof, we dualize the
points of $P$ to planes, and denote the resulting set of planes by
$P^*$.

Choose a random sample $N_0$ of $r=\lceil 2/\eps\rceil$ points
of $P$, pass to the subset $N_0^*\subseteq P^*$ of dual planes,
construct their lower envelope $E_0$, and triangulate each of its
faces, into a total of $O(r)$ triangles. We extend each of these
triangles $\tau_0$ downwards to the semi-unbounded vertical prism $\tau$
bounded from above by $\tau_0$. We thus obtain a decomposition
of the region $E_0^-$ below $E_0$ into $O(r)$ such prisms.

We take $N_0$ to be part of the output $\eps$-net.  By construction,
any lower halfspace $h$ whose bounding plane is dualized into a point
$h^*$ which lies above $E_0$ has a plane of $N_0^*$ passing below it,
so, in primal space, $h$ contains the point of $N_0$ dual to that
plane.  Thus it remains to consider halfspaces whose dual points lie
below $E_0$.

By the $\eps$-net theory, our sample $N_0$ satisfies, with high
probability, the property that each of the prisms $\tau$ in the
decomposition of $E_0^-$ is crossed by at most $\frac{cn}{r}\log r$
planes of $P^*$, for some absolute constant $c$.
We may assume that our sample does indeed satisfy this property.

Let $\tau$ be one of the prisms, and suppose that $\tau$ is crossed
by $tn/r$ planes of $P^*$. If $t< 1$, we leave $\tau$ intact.
Otherwise, we take a random sample $N^*_\tau$ of $c't\log t$ planes
from those crossing $\tau$, for some sufficiently large but absolute
constant $c'$. We may assume that each of these samples is a
$(1/t)$-net, for the corresponding value of $t$, for the range space
of the planes of $P^*$ crossing $\tau$, where each range is the subset
of planes stabbed by some downward-directed vertical ray (that is, lying below some point).

We take $N$ to be the union of $N_0$ and of all the primal sets
$N_\tau$ (dual to the corresponding $N^*_\tau$),
and claim that $N$ is an $\eps$-net. That
is, if $h$ is a lower halfspace that contains at least $\eps n$ points
of $P$ then $h$ contains a point of $N$. Indeed, as already noted,
it suffices to consider the case where the point $h^*$ dual to the
plane bounding $h$ lies in $E_0^-$. By the same reasoning, if $h^*$
lies in some initial prism $\tau$ and above the upper envelope
$E_\tau$ of $N^*_\tau$, we are also done. The remaining case is
impossible, because then the downward-directed vertical ray emanating
from $h^*$ does not meet any plane of $N^*_\tau$, and therefore meets
fewer than $\eps n$ planes of $P^*$, contradicting the assumption that
$h$ is ``heavy''.

It remains to show that $|N|=O(r)$. This follows from the exponential
decay lemma, initially established by Chazelle and
Friedman~\cite{cf-dvrsi-90}, and later extended by Agarwal
\etal~\cite{ams-cmfal-98}.  (This lemma holds with high probability
for the initial sample $N_0$, and, as above, we may assume that the
lemma does indeed hold for $N_0$.)

This completes the fifth proof of the theorem.
$\Box$

\medskip

\noindent{\bf Remarks.}
(1) It is interesting to compare the construction used in the first
four proofs to the one used in the fifth. A common feature of both
constructions is that their last step involves the construction of
$\eps'$-nets, for $\eps'$ a {\em constant},\footnote{%
  Well, in the fifth proof, constant on average.}
for many subsets of $P$.
However, these subsets are obtained by what seems to be totally
unrelated methods. It would be interesting to ``see through'' these
constructions, and perhaps show that they are more related than what
meets the eye. 

\medskip

\noindent
(2) It is worth mentioning that the construction of Pyrga and Ray, on
which our proofs are based, might be very inefficient in some simple
geometric settings (other than the one considered here). For example,
consider the range space defined by points and lines in the plane.  We
use the construction of Elekes \cite{e-spnta-02} of $n$ points and $n$
lines in the plane with $\Omega(n^{4/3})$ incidences.  Let $k$ be an
integer and put $n=2k^3$. Consider the set $P$ consisting of the
points in the integer grid $[1:k]\times [1:2k^2]$.  Put $\eps =
\frac{1}{2k^2}$. Notice that each of the lines of the form $y = ax+b$,
for $a \in [1:k]$ and $b \in [1:k^2]$ contains exactly $k = \eps n$
points of $P$. Moreover, every pair of such lines intersect in at most
one point of $P$. Hence, the collection $\F$ of these lines satisfies
conditions (a) and (b) of the construction, but $|\F|$ is way too
large, namely $|\F| = k^3= \Omega(\frac{1}{\eps^{3/2}})$.


\subsection{Deterministic construction}

In this final subsection, we note that, for the case of halfspaces in
three (and two) dimensions, one can construct linear-size $\eps$-nets
in an efficient deterministic manner, by a careful implementation of
the construction given in the first four proofs of
Theorem~\ref{main3d}, using standard techniques due to \Matousek and
others. Specifically, we will only consider the problem in three
dimensions, since the two-dimensional case can be handled as a special
case. Moreover, a different deterministic construction in the plane
follows from the technique of \Matousek in \cite{m-cen-90}.

We proceed as follows. Consider the dual version, in which we have a
collection $H$ of $n$ planes in $\reals^3$, and the ranges correspond
to points, so that the range of a point $q$ is the set of planes
passing below $q$. Consider a modified range space, in which the
ranges are defined by segments, so that the range of segment $e$ is
the set of planes of $H$ which cross $e$.

Set $\eps'=\eps/4$. We first construct an \emph{$\eps'$-approximation}
$X$, of size $|X| =O(1/\eps^2 \log(1/\eps))$, for the modified range
space.  That is, for each segment $e$, if we set $H_e$ (resp., $X_e$)
to be the set of those planes in $H$ (resp., $X$) that cross $e$, then
we have
$$
\left| \frac{|X_e|}{|X|} - \frac{|H_e|}{|H|} \right| <
\frac{\eps}{4}.
$$
One can construct $X$ deterministically in time $O(n/\eps^c)$, where
$c$ is some (small) constant, using the general technique of
\Matousek~\cite{m-dcg-96}.

Next, we construct the arrangement $\A(X)$, and extract from it
the set $S$ of all the vertices at level at most $\frac32\eps |X|$.
Our goal is to compute a maximal subset $\F\subseteq S$, such
that the crossing distance between any pair of vertices in $\F$
is at least $r := \frac14 \eps |X|$.
We do this by explicitly computing the crossing distance
(in $\A(S)$) between each pair of points of $S$, and then by
augmenting $\F$ incrementally. We first set $\F$ to be the empty set.
At each step we take the next point $z$ in $S$, and check whether it
lies at crossing distance larger than $r$ from all the points
currently in $\F$. If so, we add it to $\F$. It is easy to see
(using the fourth proof of Theorem~\ref{main3d}) that
$|\F| =O(1/\eps)$.

Next, for each point $p \in \F$, we compute a $\beta$-net $N_p$, for
some $\beta < 1/8$, for the set $H_p$ of all the planes of $H$ lying
below $p$. This takes $O(\eps n/\beta^c)$ time per point, using the
algorithm in \cite{m-dcg-96}, for an overall time $O(n/\beta^c)=O(n)$.
We take $N$ to be the union of all the sets $N_p$, and claim that (the
primal image of) $N$ is indeed an $\eps$-net for $(P,\CalH^-)$. Since
$|N| = O(|\F|) = O(1/\eps)$, we have indeed constructed an $\eps$-net
for $(P,\CalH^-)$ of size $O(1/\eps)$.

The argument that $N$ is an $\eps$-net proceeds as follows.
Denote the level of a point $u$ in $\A(H)$ by $\lambda_u$.
let $q$ be a point at level $\lambda_q=\eps n$ in $\A(H)$.
Letting $e$ denote the downward-directed ray from $q$, we have
$|H_e|=\eps n$, and therefore
$$
\frac{3\eps}{4} < \frac{|X_e|}{|X|} < \frac{5\eps}{4} .
$$
Thus $q$ lies at level between $\frac34\eps|X|$ and $\frac54\eps|X|$
in $\A(X)$. Let $q'$ be a vertex of the cell in $\A(X)$ containing
$q$. Then $q'$ lies at the same level (with respect to $X$), and in particular $q'\in S$.
This implies, as above, that $\lambda_{q'} \le \frac32\eps n$.
Since $|X_{qq'}|=0$, we have $|H_{qq'}| < \frac14\eps n$. Hence at
least $\frac34\eps n$ of the planes of $H$ passing below $q$ also pass
below $q'$. In other words, at least half of the $\lambda_{q'}$ planes
of $H$ passing below $q'$ also pass below $q$.

Now if $q'\in\F$, then the $\beta$-net $N_{q'}$ contains a plane
passing below $q$ (by our choice, $\beta<1/2$), and we are done.
Otherwise, there exists a point $q''\in\F$ such that the crossing
distance $|X_{q'q''}|$ satisfies $|X_{q'q''}| \le \frac14\eps|X|$,
so $|H_{q'q''}| \le \frac12\eps n$, and therefore
$$
|H_{qq''}| \le |H_{qq'}| + |H_{q'q''}| \le \frac34\eps n .
$$
Hence, at least $\frac14\eps n$ of the planes of $H$ below $q$ also
pass below $q''$. Observe that the level of $q''$ in $\A(H)$ is at
most $(\frac32+\frac14)\eps n < 2\eps n$, so
at least $1/8$ of the planes below $q''$ also
pass below $q$, so one of the planes of $N_{q''}$ must pass below $q$.

The resulting algorithm runs in (deterministic) time
$O(n/\eps^{c})$, where $c$ is some small constant. In this sketchy
solution we made no attempt to optimize the dependence of the running
time on $\eps$, which can probably be improved using
efficient algorithms for constructing cuttings \cite{m-ept-92}.

 \providecommand{\CNFX}[1]{ {\em{\textrm{(#1)}}}}
  \providecommand{\CNFSoCG}{\CNFX{SoCG}}  

\providecommand{\Erdos}{Erd{\H o}s\xspace}




\end{document}